\documentclass[aps,prfluids,preprint,superscriptaddress]{revtex4-1}
\usepackage{graphicx}
\usepackage{color}
\usepackage{upgreek}
\newcommand\Ja{\mbox{\textit{Ja}}}
\newcommand\Sh{\mbox{\textit{Sh}}}
\newcommand\Pe{\mbox{\textit{Pe}}}
\newcommand\Rey{\mbox{\textit{Re}}}

\begin{document}

%\preprint{}

\title{Ultrasound-enhanced mass transfer during single-bubble diffusive growth}

\author{\'Alvaro Moreno Soto}
\email{amorenos@mit.edu}
\affiliation{Physics of Fluids group, Max-Planck Center Twente for Complex Fluid Dynamics, Department of Science and Technology, Mesa+ Institute, and J. M. Burgers Center for Fluid Dynamics, University of Twente, P.O. Box 217, 7500 AE Enschede, The Netherlands}
\affiliation{Department of Mechanical Engineering, Massachusetts Institute of Technology, 77 Massachusetts Avenue, Cambridge, Massachusetts, 02139, USA}

\author{Pablo Pe\~nas}
\email{p.penaslopez@utwente.nl}
\author{Guillaume Lajoinie}
\author{Detlef Lohse}
\author{Devaraj van der Meer}
\affiliation{Physics of Fluids group, Max-Planck Center Twente for Complex Fluid Dynamics, Department of Science and Technology, Mesa+ Institute, and J. M. Burgers Center for Fluid Dynamics, University of Twente, P.O. Box 217, 7500 AE Enschede, The Netherlands}

\date{\today}

\begin{abstract}
Ultrasound is known to enhance surface bubble growth and removal in catalytic and microfluidic applications, yet the contributions of rectified diffusion and microstreaming phenomena towards mass transfer remain unclear.
We quantify the effect of ultrasound on the diffusive growth of a single  spherical CO$_2$ bubble growing on a substrate in supersaturated water. 
The time dependent bubble size, shape, oscillation amplitude and microstreaming flow field are resolved.  
We show and explain how ultrasound can enhance the diffusive growth of surface bubbles by up to two orders of magnitude during volumetric resonance. The proximity of the wall forces the bubble to oscillate non-spherically, thereby generating vigorous streaming during resonance that results in convection-dominated growth.

\end{abstract}

%\keywords{}

\maketitle

%%%%%%%%%%%%%%%%%%%%%%%%%%%%%%%%%%%%
%%% INTRODUCTION
%%%%%%%%%%%%%%%%%%%%%%%%%%%%%%%%%%%%
\section{Introduction}
%The formation of bubbles on the surface of gas-evolving catalysts can be detrimental towards the energy efficiency of many electrochemical systems \cite{Zhao2019}. 
%The use of ultrasound is a promising technique to enhance catalytic bubble removal \cite{Taqieddin2017}, whereas in medicine, ultrasound-driven microbubbles have widely established applications in ultrasound imaging \cite{Coussios2008} and targeted drug delivery \cite{deCock2016, Lajoinie2018}. 
Ultrasound application is a promising intensification technology with the ability to improve the energy efficiency of electrochemical reactions by promoting bubble detachment from the catalyst surface \cite{Li2009},  accelerate liquid degassing through cavitation \cite{Kapustina1965} or enhance mass transfer processes  in gas--liquid micro-sono-reactors \cite{Dong2015}. 
From a detrimental aspect,   gas diffusion across ultrasound-driven microbubbles employed in biomedical acoustic therapies and diagnostics  \cite{Coussios2008} may substantially alter the bubble size \cite{OBrien2013} or longevity \cite{Bader2018}. Similarly, oscillating bubbles driving microfluidic applications \cite{Hashmi2012} or sonochemical reactions \cite{FernandezRivas2012} are generally surrounded by non-degassed liquids and unwanted mass transfer effects may become significant during continued ultrasonic operation \cite{Iida2007}.

A gas bubble undergoing volume oscillations in a liquid--gas solution experiences a mass transfer enhancement that is believed to result from two phenomena. The first is rectified diffusion \cite{Eller1965}, consequence of the asymmetries in the mass transfer rate across the bubble during the  expansion and compression half-cycles, generally favoring growth. 
The second is acoustic microstreaming \cite{Elder1959}, a second-order (in driving amplitude) steady flow driven by non-spherical bubble oscillations \cite{Marmottant2006}.
Microstreaming essentially renews the gas content of the liquid in contact with the bubble \cite{Church1988}. 

The ultrasound-enhanced growth of gas bubbles attached to surfaces remains a poorly studied subject. Surface bubbles always oscillate non-spherically, unavoidably giving rise to microstreaming. Consequently, the classical theories of rectified diffusion \cite{Crum1984, Fyrillas1994, Brenner2002} are no longer applicable, and the mass transfer process remains unclear. In this work, we conduct unprecedented experiments that quantify the effect of ultrasound on the diffusive growth of a single monocomponent surface bubble in supersaturated water. To fully capture the physics, we resolve both bubble dynamics and streaming flow field as the bubble overgrows its resonant size.

The diffusive growth rate of an unperturbed bubble is best quantified by the Jakob number for mass diffusion \cite{Szekely1971, Prosperetti2017},
$\Ja = (C_\infty-C_s)/\rho_g$,
where $C_\infty$ is the mass concentration of dissolved gas in the ambient liquid, $C_s$ the saturation mass concentration at the bubble surface, and $\rho_g$ the density of gas in the bubble.  For small  Laplace pressures, $\Ja$ can be assumed independent of the bubble size; if so, $\Ja$ strictly represents the product of the degree of supersaturation ($C_\infty/C_s-1$, the driving force) and the dimensionless Henry solubility ($C_s/\rho_g$, the growth-rate amplifier). 

Previous experimental studies are mostly constrained to 
isolated air (multicomponent) bubbles in water under (close to) saturation conditions ($C_\infty/C_s \approx 1$, $\Ja \approx 0$) \cite{Eller1969, Gould1974, Crum1980, Lee2005, Leong2010, Leong2011}. 
Such bubbles slowly dissolve due to surface tension, unless sonicated above a certain threshold amplitude. Kapustina \cite{Kapustina1973} exceptionally studied the growth of air bubbles on a needle in water notably supersaturated with air ($C_\infty/C_s \sim 1.5$, $\Ja < 10^{-2}$). In contrast, our experiments are performed at larger $\Ja \sim 0.1$. All studies \cite{Eller1969, Gould1974, Crum1980, Lee2005, Leong2010, Leong2011, Kapustina1973} coincide in that the appearance of shape oscillations is often accompanied by vigorous streaming and enhanced rates of mass transfer.

%%%%%%%%%%%%%%%%%%%%%%%%%%%%%%%%%%%%
%%% EXPERIMENTAL
%%%%%%%%%%%%%%%%%%%%%%%%%%%%%%%%%%%%
\section{Experimental set-up}
\begin{figure}[h!]
 \includegraphics[width=0.8\columnwidth]{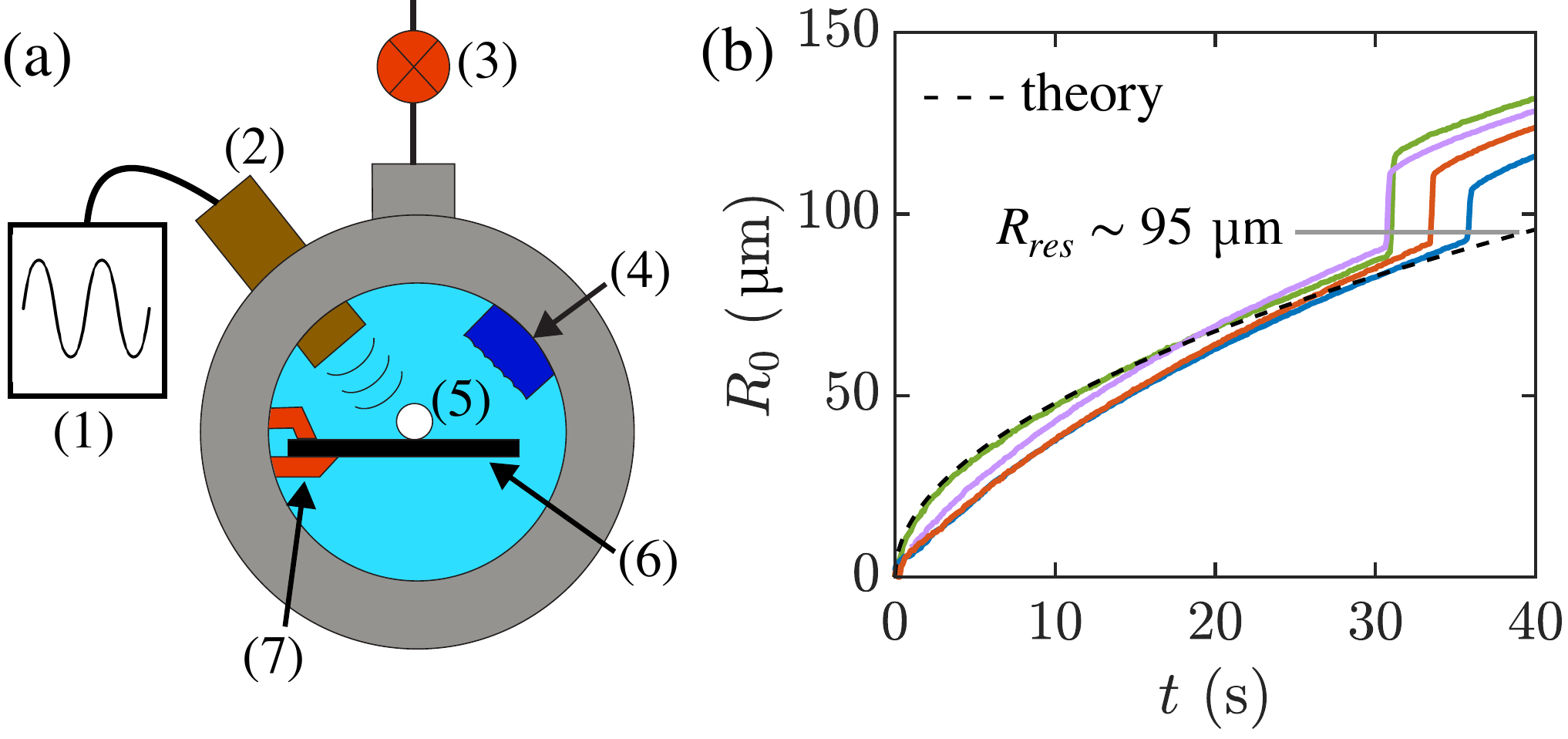}%
 \caption{\label{fig1} (a) Experimental set-up (camera view). (1) Waveform generator, (2) ultrasonic transducer, (3) pressure valve, (4) acoustic absorber, (5) CO$_2$ bubble, (6) silicon substrate, (7) holder.
 (b) Evolution of the mean bubble radius in time for four bubbles subjected to different acoustic pressure amplitudes. The dashed line is the theoretical prediction for purely diffusive growth on a substrate, namely Eq. (\ref{eq:EP}) with $\Ja = 0.093$ as a free parameter.}
 \end{figure}

 The experimental set-up is sketched in Fig. \ref{fig1}(a). A pressurized tank ($P_\mathit{sat} \approx 4$ bar) is first filled with carbonated water saturated at the same pressure, which is subsequently lowered to $P_0 \approx 3.5$ bar. Consequently, the solution becomes supersaturated ($C_\infty/C_s = P_\mathit{sat} / P_0 \approx  1.14$, nominal $\Ja = 0.13$) and a single CO$_2$ bubble spontaneously nucleates and grows from a hydrophobic cavity (20 $\upmu$m diameter) etched on an otherwise hydrophilic silicon substrate. The bubble keeps growing until detachment.
Meanwhile, a transducer (Benthowave BII-7501/50) constantly generates ultrasound waves of frequency $\omega/2\pi =  50$ kHz. The pressure amplitude was kept constant throughout the lifetime of a given bubble but was varied (nominal values ranging 2--12 kPa) between experiments. Unfortunately, the exact driving amplitude transmitted to the bubble remained uncertain since it was uncontrollably weakened by a substantial number of parasitic bubbles that formed on the transducer surface, in addition to the likely formation of standing waves within the experimental tank.

%%%%%%%%%%%%%%%%%%%%%%%%%%%%%%%%%%%%
%%% FULL GROWTH DYNAMICS
%%%%%%%%%%%%%%%%%%%%%%%%%%%%%%%%%%%%
\section{Growth dynamics}
The bubble-growth process was captured by a high-speed camera (Photron SA-Z) at 1000 fps.
As seen in Fig. \ref{fig1}(b), the ambient bubble radius $R_0$ (defined as the mean sphere-equivalent radius about which the bubble oscillates) grows by diffusion until it approaches the resonant size. The early growth dynamics are well predicted by the classical Epstein--Plesset theory for diffusive growth \cite{Epstein1950} despite some deviations which can be attributed to perturbations in the initial condition of the concentration field \cite{Penas2017}. Taking into account the presence of the substrate, the asymptotic solution reads \cite{Enriquez2014}
\begin{equation}\label{eq:EP}
\frac{R_0^2}{D t} = 2 \Ja\left[
\left(\frac{\Ja}{2\pi}\right)^{1/2} + \left(\frac{1}{2} + \frac{\Ja}{2\pi}\right)^{1/2} \right]^2,
\end{equation}
where $D = 1.76 \times 10^{-9}$ m$^2$/s is the mass diffusivity of CO$_2$ in water. The effect of surface tension (crucial in equilibrated solutions) has been neglected by virtue of the relatively strong supersaturation. The value of $\Ja = 0.093$ used in Fig. \ref{fig1}(b) is smaller than the nominal value, consequence of the considerable degassing effect \cite{Kapustina1965, Kapustina1973, Moreno2017} of continuous ultrasonic operation.

During resonance, the growth rate deviates and strikingly increases by up to two orders of magnitude. Once the bubble outgrows resonance, it continues growing diffusively until detaching at $R_0 \approx 330$ $\upmu$m \cite{Moreno2017}. 
Immediate detachment during resonance was otherwise observed for higher acoustic amplitudes.

Ignoring the effect of surface tension, the natural frequency of a spherical surface bubble is given by \cite{Overvelde2010, Xi2014}
\begin{equation}
\omega_0 \approx \sqrt{\frac{2}{3} \left( \frac{3 \kappa P_0}{\rho_l R_0^2}\right)}.
\end{equation}
One may check that $\omega_0$ is precisely equal to the driving frequency $\omega$ when the bubble attains the resonant size of $R_0 \approx 90$--95 $\upmu$m. Here, $\rho_l$ is the water density and $\kappa$ is the polytropic exponent. The bubbles are expectedly adiabatic at resonance ($\kappa \approx 1.28$), given that $R_0/(\alpha_g/\omega)^{1/2}\sim 25 \gg 1 $ and $R_0/(\alpha_l/\omega)^{1/2}\sim 130 \gg 1 $ \cite{Prosperetti1977}, where $\alpha_g$ and $\alpha_l$ denote the thermal diffusivities of the gas inside the bubble (CO$_2$) and the surrounding liquid (water) respectively. 
Neglecting surface tension $\sigma$ on $\omega_0$ is justified since $2\sigma/R_0P_0 \sim 0.01 \ll 1$ when resonance occurs.

%%%%%%%%%%%%%%%%%%%%%%%%%%%%%%%%%%%%
%%% OSCILLATION DYNAMICS
%%%%%%%%%%%%%%%%%%%%%%%%%%%%%%%%%%%%
\begin{figure}[h!]
\includegraphics[width=0.7\columnwidth]{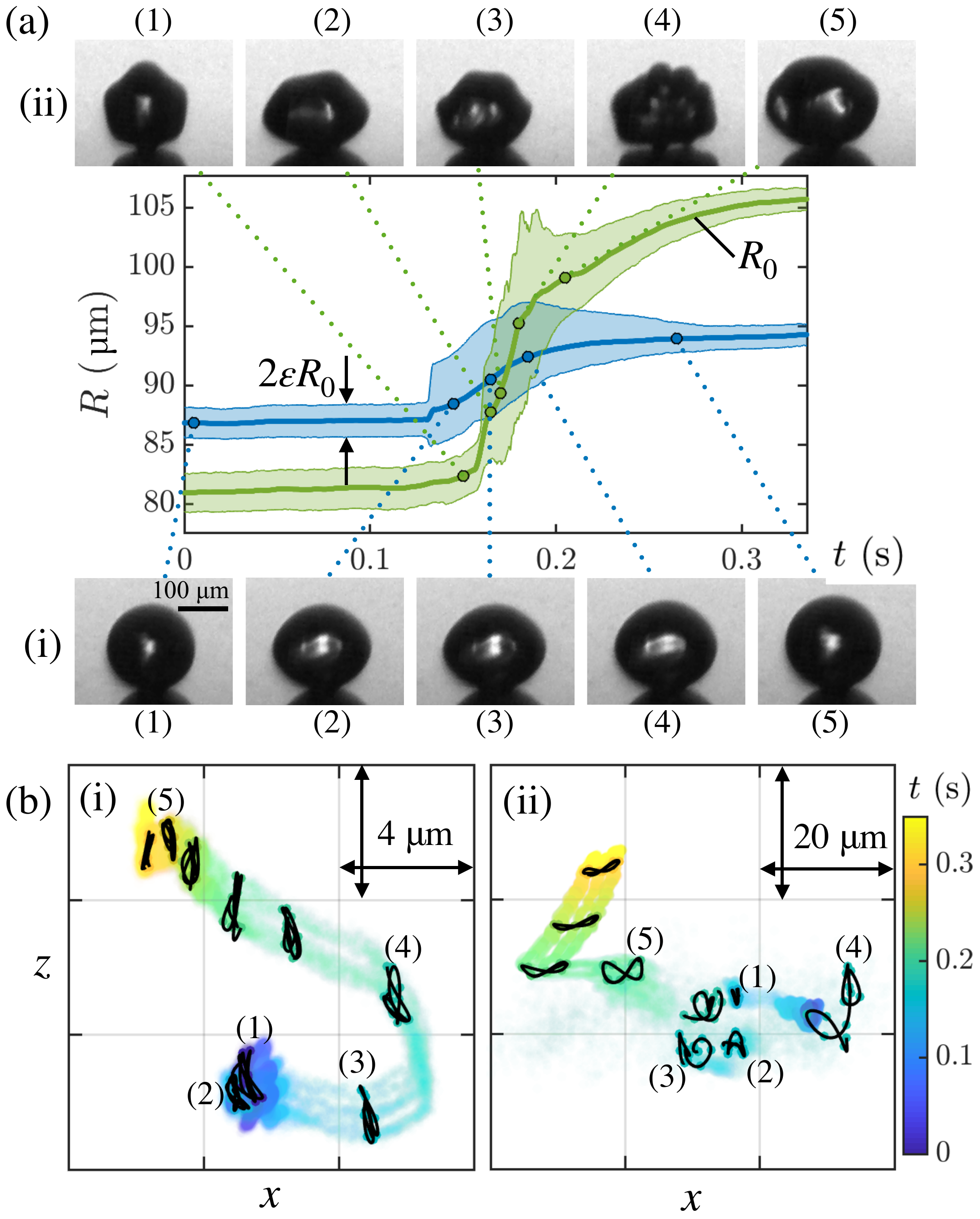}%
\caption{\label{fig2} (a) Radius growth dynamics of two bubbles during resonance exposed to a comparatively (i) low and (ii) high amplitude of acoustic pressure; nominal values are approximately (i) 2 kPa and (ii) 10 kPa. The real pressure delivered remains unknown and may be quite smaller than the nominal value.  The time origin is arbitrary. The continuous line represents the mean sphere-equivalent radius $R_0$; the envelope delimits the volumetric oscillation amplitude around $R_0$. (b) Position map of the center of mass for bubbles (i) and (ii) in the vertical plane as a function of time. Solid  lines represent the signature of the centroid position over three cycles (0.06 ms) at different times. Signatures labeled with numbers (1--5) correspond in time with the bubble side-view snapshots. The color code represents the evolution in time.}
 \end{figure}

To obtain more insight into the bubble dynamics, we record the bubble oscillating at resonance at 200,000 fps (4 frames per oscillation cycle).
Owing to the proximity of the wall, the volumetric oscillations  of the bubble occur concurrently with translational oscillations of its center of mass \cite{Marmottant2006b}.  At low acoustic amplitudes or sufficiently far from resonance, perturbations are weak and follow the ``Narcissus" effect \cite{Marmottant2006b}: Translational oscillations occur mainly in the vertical direction perpendicular to the wall [Fig. \ref{fig2}(b), panel (i)].  The snapshots of bubble (i) in Fig. \ref{fig2}(a) evidence the oblate shape that the bubble assumes during the expansion %(low pressure) 
half-cycle \cite{Lajoinie2018}. 

Stronger driving pressure amplitudes do not only increase the volumetric oscillation amplitudes, but also lead to a rich variety of surface modes [Fig. \ref{fig2}(a), bubble (ii)]. The latter occur only during resonance, since the threshold amplitude for shape instability is precisely minimal at the volumetric resonance size \cite{Versluis2010}. Lamb's classical expression \cite{Lamb1895, Versluis2010} predicts the most unstable surface mode at $R_0 = 95$ $\upmu$m to be $n=5$, in agreement with our experimental observations [snapshot (ii-1)]. Furthermore, the center of mass undergoes extensive translations, presumably in all directions \cite{Tho2007}. It is seen that the center-of-mass oscillations [Fig. \ref{fig2}(b), panel (ii)], albeit heterogeneous in time, have a predominant horizontal component. A $\infty$-shaped signature was commonly found.

The (dimensionless) volumetric oscillation amplitudes normalized by $R_0$ at the moment of maximum growth are $\varepsilon < 0.08$, i.e., four times larger, at best, than those of the linear oscillations observed immediately before or after resonance (typically $\varepsilon \sim 0.02$), whereas the growth rates can easily differ by two orders of magnitude. 
It is evident from Fig. \ref{fig2}(a) that a larger $\varepsilon$ and the onset of surface modes are clearly associated with a faster bubble growth rate ($\dot R_0$).
However, these findings cannot be explained by rectified diffusion alone: Substituting these values of $\varepsilon$ into a model adapted from \citet{Crum1984} results in an underestimation of the order of magnitude of the observed maximum growth rate (see Supplemental Material).
According to the model, the increment in the growth rate that rectification offers  is small relative to the already large diffusive growth rate. 
Rectified diffusion therefore remains a non-critical and subdominant mechanism to both (i) the unperturbed diffusive mass transfer away from resonance, and (ii) the microstreaming-enhanced mass transfer during resonance. 
We stress that condition (i) applies by virtue of the strong level of supersaturation in the liquid: The rectified diffusion enhancement offered by the modest acoustic amplitudes applied is small compared to the fast diffusive growth. In contrast, in a gas-equilibrated solution (i.e., at a very low super- or under-saturation) the same amount of mass rectification would become critical and comparatively very significant. Note that the horizontal translations suffered by the bubble during resonance, while important, also remain a second order factor in the total mass transfer increase as compared to the effect of microstreaming.

%%%%%%%%%%%%%%%%%%%%%%%%%%%%%%%%%%%%
%%% MICROSTREAMING
%%%%%%%%%%%%%%%%%%%%%%%%%%%%%%%%%%%%
\section{Microstreaming}
\begin{figure*}[ht!]
 \includegraphics[width=1.0\textwidth]{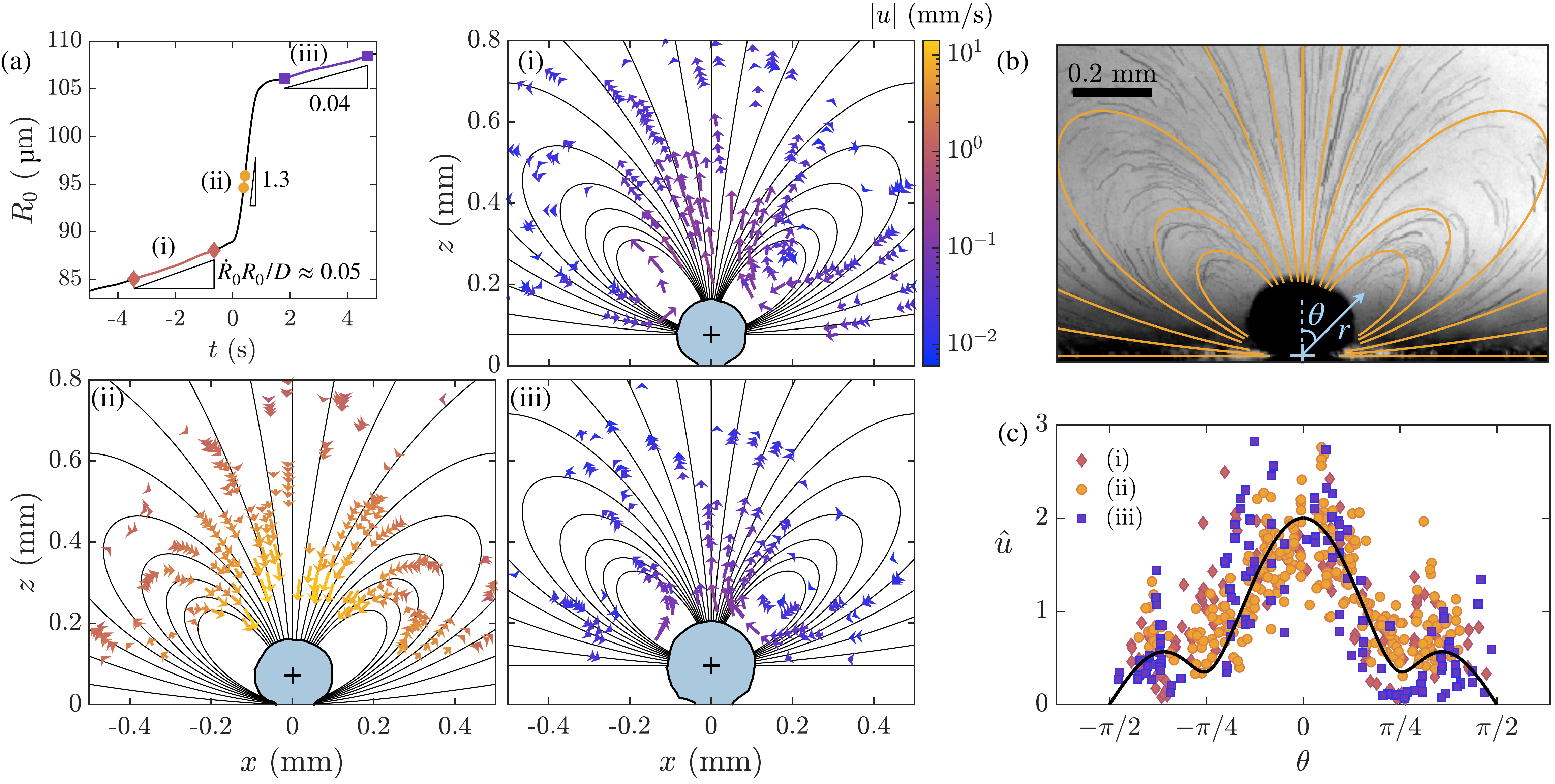}%
 \caption{\label{fig3} 
 (a) The first panel shows a typical evolution of the mean bubble radius during resonance (time origin is arbitrary). The nominal acoustic pressure amplitude is approximately 10 kPa. The arrows in the three remaining panels (i--iii) correspond to particle velocities measured during three distinct time phases highlighted along the radius growth curve. 
 The black lines are the theoretical far-field streamlines according to Eq. (\ref{eq:streamfunction}); in (b) these are directly superimposed on the particle pathlines covered during phase (ii) which spans 0.25 s. 
 (c) Rescaled far-field velocity magnitude $\hat u = |u/A|(r/R_0)^3/R_0\omega$ as a function of $\theta$ for phases (i--iii). The characteristic source strength $A$ is computed by fitting each experimental velocity set to the theoretical curve (solid black line) given by Eq. (\ref{angvel}). We obtain (i) $A = 7.5 \times 10^{-5}$, (ii) $A = -1 \times 10^{-2}$ and (iii) $A = 4 \times 10^{-5}$. 
 See also supplementary movie. }
 \end{figure*}

Microstreaming must therefore be the leading-order contributor to the magnitude of maximum $\dot R_0$ during resonance. 
This hypothesis was verified in further experiments where 0.43 ml/l of neutrally buoyant 3-$\upmu$m polystyrene latex beads (Sigma-Aldrich) were added to the solution. The streaming flow field generated as the bubble grows through resonance was visualized by means of particle tracking velocimetry \cite{Bolanos2017}; images were recorded at 1000 fps. The particle velocities measured during three distinct phases for a particular experiment (provided as a supplementary movie) are shown in Fig. \ref{fig3}. In each of the fifteen streaming experiments that were conducted, a qualitatively identical streaming behavior was observed.

Immediately before resonance [phase (i)], the bubble undergoes weak volumetric and vertical translation oscillations [cf. Fig. \ref{fig2}, bubble (i)].  We observe weak `fountain mode' streaming [Fig. \ref{fig3}(a), panel (i)], generically reported in similar microbubble streaming experiments \cite{Marmottant2003, Marmottant2006, Marmottant2006b, Lajoinie2018, Elder1959, Bolanos2017}.
The streaming velocity close to the bubble, $U_s$, is approximately 0.2 mm/s and the streaming Reynolds number is $\Rey = 2R_0 U_s/\nu \approx 0.04$, where $\nu$ is the kinematic viscosity of the liquid.
The P\'eclet number based on $U_s$ remains fairly small at $\Pe = 2R_0 U_s/D \approx  28$.
During resonance [phase (ii)], the bubble undergoes strong volumetric, translational and surface oscillations [cf. Fig. \ref{fig2}, bubble (ii)]. 
Strikingly, the direction of streaming reverses (`antifountain' mode) [Fig. \ref{fig3}(a), panel (ii)]. The streaming velocities then escalate by two orders of magnitude ($U_s \approx 15$ mm/s, $\Rey \approx 3$, $\Pe \approx  2100$), and so does the growth rate. There is a second reversal in direction immediately after resonance and all the attributes prior to resonance are recovered [phase (iii)].  Weak fountain-mode streaming ($U_s \approx 0.18$ mm/s, $\Rey \approx 0.04$, $\Pe \approx  21$) is observed once again.

The onset of surface mode activity is known to induce notably vigorous streaming \cite{Tho2007, Gould1974}, whereas a similar reversal from fountain to antifountain mode was first encountered by \citet{Elder1959}, which he attributed to the onset of higher-order surface modes triggered at sufficiently large acoustic amplitudes. Streaming patterns and direction are indeed dictated by the modes of microbubble oscillation \cite{Tho2007, Wang2013, Rallabandi2014, Cleve2019}, which are frequency and amplitude dependent. 

The particle pathlines sufficiently far from the bubble were found to be well described by the streamlines corresponding to the  leading-order far-field (dipole-like) axisymmetric streamfunction proposed by \citet{Marmottant2003},
 \begin{equation}\label{eq:streamfunction}
\Psi(r, \ \theta) = \frac{A R_0^4\omega}{r}\cos^2 \theta \sin^2 \theta,
\end{equation}
where $A$ is the dimensionless source strength and $r$, $\theta$ are the spherical coordinates defined in Fig. \ref{fig3}(b).
The best agreement resulted after relocating the coordinate origin of the theoretical streamfunction on the bubble's center of mass for the fountain mode [Fig. \ref{fig3}(a) panels (i, iii)], and on the bubble base for the antifountain mode [Fig. \ref{fig3}(a) panel (ii), Fig. \ref{fig3}(b)]. A likely explanation is the possible existence of a weak recirculation zone very close the wall \cite{Rallabandi2014} (here unobservable)  that arises during fountain-mode streaming only [phases (i) and (iii)].

The source strength $A$ depends on the oscillation mode of the bubble and is a priori unknown. Typically $A$ is $O(\varepsilon^2)$ \cite{Marmottant2003, Marmottant2006, Marmottant2006b}, where $\varepsilon$ represents some characteristic oscillation amplitude. 
However, $A$ may be estimated directly from the far-field velocities \cite{Bolanos2017}. It follows that the velocity magnitude $|u|$ is  a function of $\theta$ only when rescaled in the following manner:
\begin{equation}\label{angvel}
\frac{|u/A|}{R_0\omega}\left(\frac{r}{R_0}\right)^{3} = \sqrt{\frac{\sin ^2(4\theta)}{4\sin^2 \theta}+\sin^2\theta \cos^4 \theta}.
\end{equation}
We find that $A$ drastically increases from $O(10^{-4})$ in phases (i, iii) to $O(10^{-2})$ in phase (ii) [see Fig \ref{fig3}(c)], following the two order of magnitude increase in the values of $Pe$. Note the sign reversal for the antifountain mode.

%%%%%%%%%%%%%%%%%%%%%%%%%%%%%%%%%%%%
%%% TRANSITION TO CONVECTIVE GROWTH
%%%%%%%%%%%%%%%%%%%%%%%%%%%%%%%%%%%%
\section{Mass transfer mechanism}
\begin{figure}[ht!]
 \includegraphics[width=0.6\columnwidth]{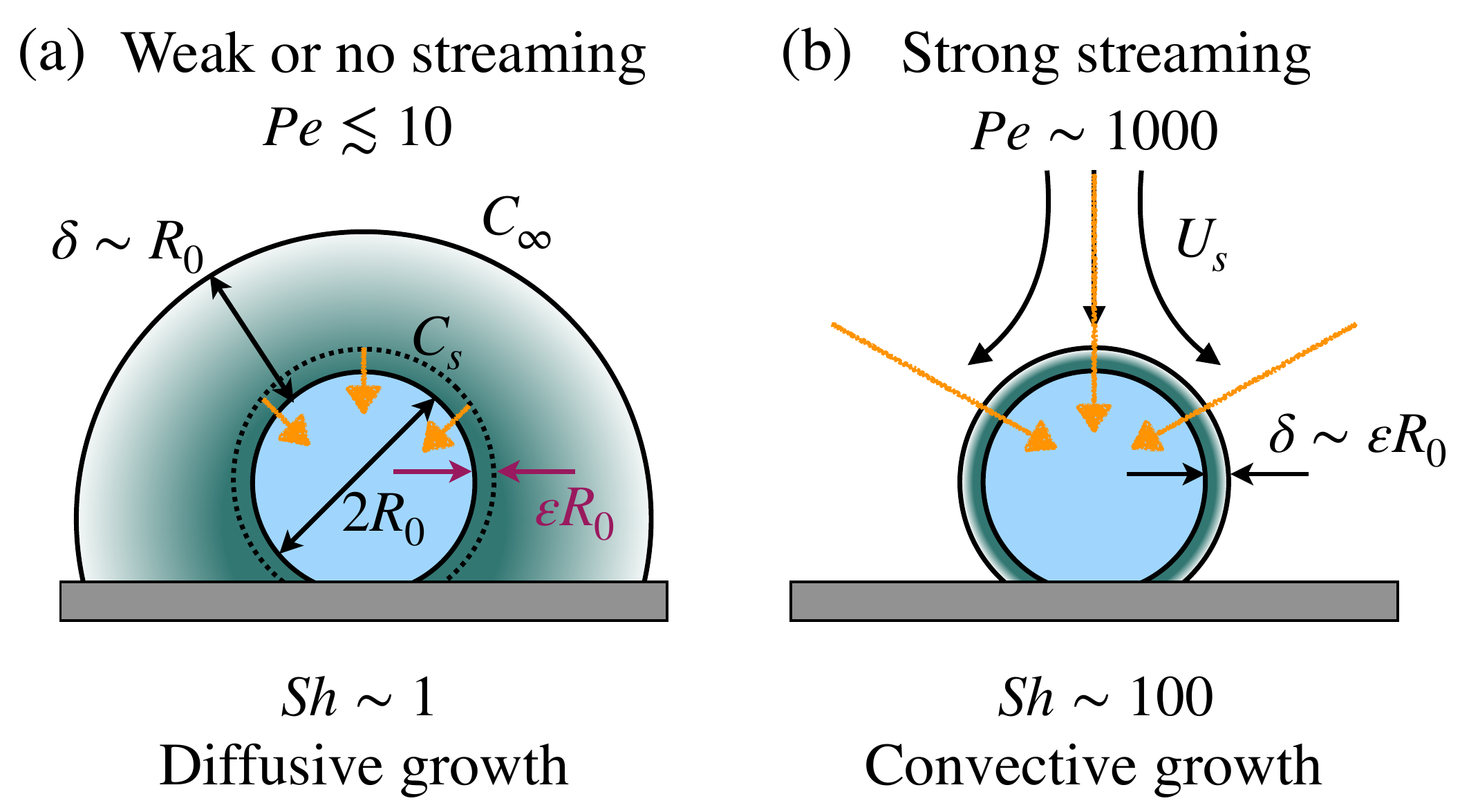}%
 \caption{\label{fig4} Schematic of the concentration boundary layer thickness $\delta$ during (a) weak streaming and (b) strong streaming. The orange arrows represent the magnitude of the mass transfer;  $\varepsilon R_0$ denotes the characteristic oscillation amplitude.}
 \end{figure}

These results indicate that, during resonance, the microstreaming velocities become large enough to induce a transition from diffusion-dominated to convection-dominated growth. 
A physical explanation can be given in terms of the diffusion layer $\delta$ (see e.g. \cite{Tobias1952}), namely the characteristic thickness of the concentration boundary layer surrounding the bubble. It follows that $\delta$ scales as $\delta/R_0 = 2/\Sh$, where the bubble Sherwood number $\Sh = (1/\Ja)(2\dot R_0 R_0/D)$ \cite{Enriquez2014, Moreno2017} constitutes a measure of the mass transfer rate. Note that $\Sh$, hence $\delta$, can be computed independently without any knowledge of the streaming velocity. 
When there is weak or no streaming [Fig. \ref{fig4}(a)] we find diffusive growth, $\Sh \sim 1$, i.e., $\delta/R_0 \sim 1 \gg \varepsilon$. The bubble oscillations only perturb the boundary layer slightly.
Strong streaming is responsible for the thinning of $\delta$ [Fig. \ref{fig4}(b)], resulting in steep interfacial concentration gradients, hence $\Sh \gg 1$. The local mixing induced by the bubble oscillatory interface is now more relevant since $\delta/R_0 \sim \varepsilon$. Nonetheless, to leading order, $\Sh$ must only depend on $\Pe$ provided that $\Pe$ is large.
Naturally, amplifying the acoustic driving strengthens microstreaming: $\Pe$  and hence $\Sh$ increase.

\begin{figure}[h]
 \includegraphics[width=0.7\columnwidth]{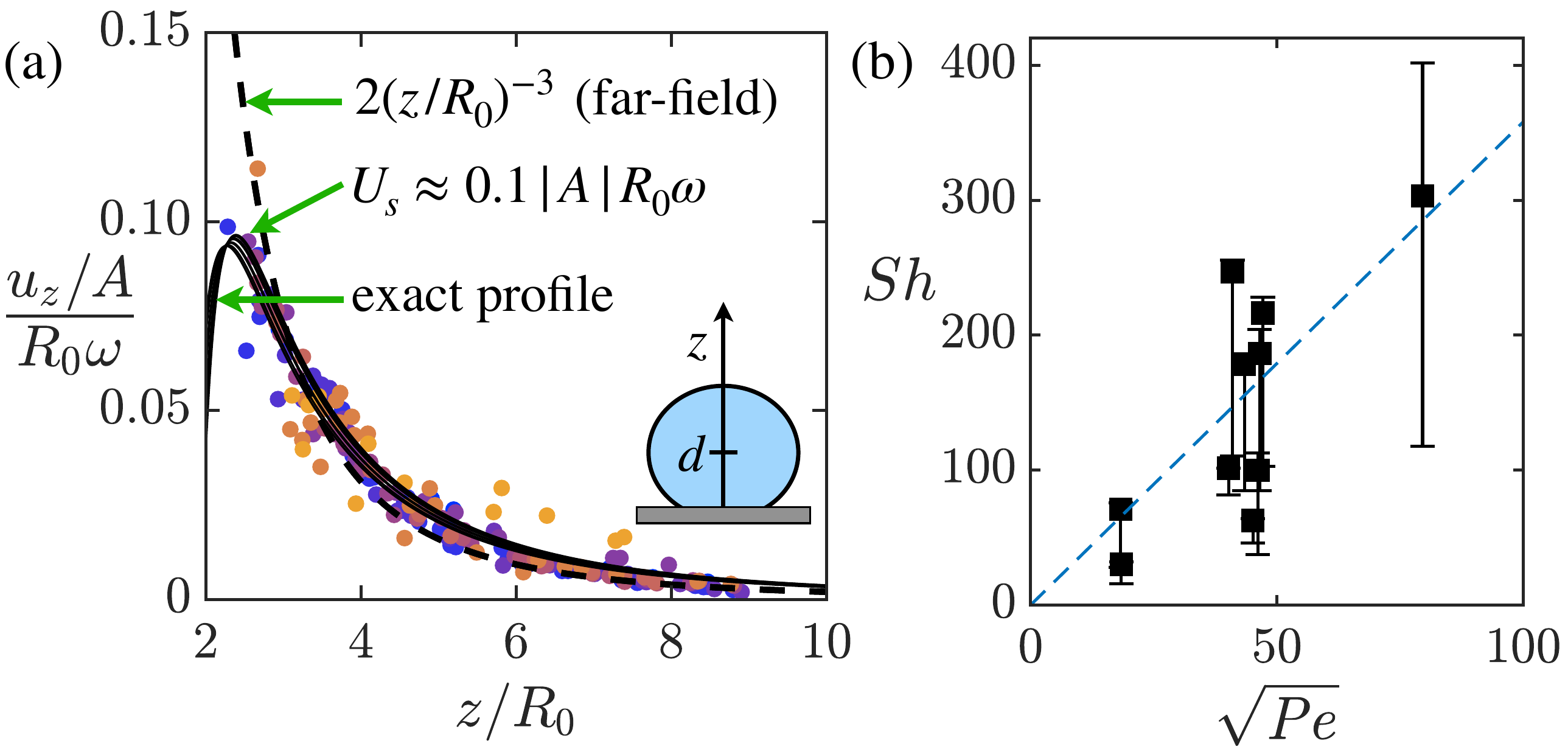}%
 \caption{\label{fig5} 
 ($a$) Vertical streaming velocity profiles induced by ten different bubbles (exposed to different acoustic amplitudes) during the moment of maximum growth. For each bubble, the measured particle velocities (markers) are all rescaled by a unique source strength $A$ (free parameter) to match the theoretical (``exact'') profile given by Eq. (\ref{eq:profile}). The maximum streaming velocity $U_s$ is then computed from the highest value of theoretical velocity. ($b$) Sherwood number of the same ten bubbles (at the moment of maximum growth) as a function of the P\'eclet number, $2R_0U_s/D$. Dashed line is a linear fit, $\Sh = 3.6\sqrt{\Pe}$. The error bars arise from the uncertainty in $\Ja$ due to the continuous degassing of the solution. 
Setting $\Ja$ equal to the nominal value yields a conservative lower bound on $\Sh$; computing $\Ja$ from Eq. (\ref{eq:EP}) assuming that the growth rate prior to resonance [cf. Fig. \ref{fig3}(a), phase (i)] is purely diffusive generally yields a $\Sh$ close to the upper bound.}
 \end{figure}

%%%%%%%%%%%%%%%%%%%%%%%%%%%%%%%%%%%%
%%% MASS TRANSFER AT HIGH PE
%%%%%%%%%%%%%%%%%%%%%%%%%%%%%%%%%%%% 
The above explanation---only applicable when rectified diffusion is subdominant---calls for a quantitative relation between $\Pe = 2R_0 U_s/D$ and $\Sh$ at the moment of maximum bubble growth. 
To obtain such a relation, the maximum streaming velocity surrounding the bubble, $U_s$, was first extracted in a consistent manner from a set of ten different experiments. Direct tracking of particle velocities adjacent to the bubble surface during strong (antifountain) streaming proved unviable due to optical limitations. This was circumvented by extrapolating the particle velocities measured along the $z$-axis as described in Fig. \ref{fig5}(a). The measured velocity profiles along the $z$-axis were all well described by the theoretical expression \cite{Lajoinie2018}
\begin{equation}\label{eq:profile}
\frac{u_z}{A R_0 \omega} = \frac{1}{3}(f_\mathit{stk}
+ f_\mathit{dip}+f_\mathit{hexdp}),
\end{equation}
where $f_\mathit{stk}$, $f_\mathit{dip}$ and $f_\mathit{hexdp}$, defined in Eqs. (8)--(10) in Ref. \cite{Lajoinie2018}, are dimensionless functions of
$z/R_0$ and $d/R_0$, where $d$ refers to the height of the bubble center. 
In the far-field limit ($z/R_0 \rightarrow \infty)$, Eq. (\ref{eq:profile})  simplifies to $u_z/A R_0 \omega = 2 (z/R_0)^{-3}$ [see Fig. \ref{fig5}(a)], consistent with Eq. (\ref{angvel}) when $\theta = 0$, $r = z$. 
Plotting Eq. (\ref{eq:profile}) reveals that the maximum streaming velocity is related to the source strength through $U_s \approx 0.1 |A| R_0 \omega$. 

Our measurements [Fig. \ref{fig5}(b)] suggest that, at the moment of maximum growth (where $\Pe$ is large), mass transfer is consistent with the scaling law $\Sh = C \sqrt{Pe}$, where $C$ is a constant of order unity. This functional dependence was in fact theoretically derived by \citet{Kapustina1968} and later \citet{Davidson1971}. It is no coincidence that the convective mass transfer of a freely rising bubble (see e.g. \cite{Clift1978, Takemura1998, Colombet2015}) follows this same relation.

%%%%%%%%%%%%%%%%%%%%%%%%%%%%%%%%%%%%
%%% CONCLUSIONS
%%%%%%%%%%%%%%%%%%%%%%%%%%%%%%%%%%%%
\section{Conclusions}
In summary, ultrasound can easily enhance the growth of surface bubbles by two orders of magnitude during volumetric resonance. The underlying physical mechanism is as follows: The proximity of the wall forces the spherical bubble to oscillate non-spherically.
Approaching resonance, the bubble undergoes small volumetric oscillations superimposed with translational oscillations perpendicular to the wall. The concomitant fountain-like streaming is weak and the bubble growth rate remains primarily driven by diffusion. 
During resonance, however, the onset of surface oscillations and larger translational oscillations lead to vigorous streaming, resulting in convective growth. Streaming is held majorly responsible for the mass transfer enhancement: Gas-rich liquid advected from the bulk disrupts the concentration boundary layer surrounding the bubble, greatly strengthening the concentration gradients therein. 
Our findings have direct impact on diverse applications concerning ultrasonic-driven spherical or quasi-spherical bubbles attached to a solid surface; namely, in microfluidic devices \cite{Hashmi2012}, sonochemical reactors \cite{FernandezRivas2012}, gas-evolving electrodes \cite{Zhao2019} or catalysts \cite{Lv2017}, or even during heterogeneous cavitation \cite{Bremond2006} and pool nucleation boiling \cite{Dhir2007} of gas-containing vapor bubbles.

%%%%%%%%%%%%%%%%%%%%%%%%%%%%%%%%%%%%
%%% MCEC
%%%%%%%%%%%%%%%%%%%%%%%%%%%%%%%%%%%%
\begin{acknowledgments}
This work was supported by the Netherlands Center for Multiscale Catalytic Energy Conversion (MCEC), an NWO Gravitation program funded by the Ministry of Education, Culture and Science of the government of the Netherlands.
\end{acknowledgments}

%%%%%%%%%%%%%%%%%%%%%%%%%%%%%%%%%%%%
%%% REFERENCES
%%%%%%%%%%%%%%%%%%%%%%%%%%%%%%%%%%%%
\bibliography{USBubbleGrowthReferencesV4}
\end{document}